\documentclass[preprint,aps,floatfix]{revtex4-1}
\usepackage{graphicx}
\newcommand{\B}{{\bf B}}
\newcommand{\E}{{\bf E}}

\newcommand{\bh}{\hat{\bf B}}

\newcommand{\br}{{\bf r}}
\newcommand{\bac}{{\bf a}}

\newcommand{\p}{{\bf p}}
\newcommand{\dt}{{\Delta t}}

\newcommand{\ta}{{\theta}}

\newcommand{\bv}{{\bf v}}

\newcommand{\w}{\omega}
\newcommand{\e}{{\rm e}}

\newcommand{\nn}{\nonumber}
\newcommand{\cc}{(\hat{\bf B}\times\ )}

\newcommand{\la}{\label} 

\newcommand{\be}{\begin{equation}}
\newcommand{\ee}{\end{equation}}
\newcommand{\ba}{\begin{eqnarray}}
\newcommand{\ea}{\end{eqnarray}}

\begin{document}
\title{A fundamental derivation of two Boris solvers and the Ge-Marsden theorem}

\author{Siu A. Chin}
\email{chin@physics.tamu.edu.}
\affiliation{Department of Physics and Astronomy, Texas A\&M University,
	College Station, TX 77843, USA}
%
\begin{abstract}	
	
For a separable Hamiltonian, there are two fundamental, time-symmetric, second-order
velocity-Verlet (VV) and position-Verlet (PV) symplectic integrators.
Similarly, there are two VV and PV version of exact energy conserving algorithms
for solving magnetic field trajectories. For a constant magnetic field,
both algorithms can be further modified so that their trajectories are exactly on the gyro-circle.
The magnetic PV integrator then becomes the well known Boris solver, while 
VV yields a second, previously unknown, Boris-type algorithm. Remarkably, the required
on-orbit modification is a reparametrization of the time step, 
reminiscent of the Ge-Marsden theorem. 

\end{abstract}
\maketitle

\section {Introduction}
\la{in}

The Ge-Marsden\cite{ge88} theorem states that if a symplectic integrator (SI) exactly conserves the energy, 
but no other constants of motion, then its trajectory must be exact, up to ``time-reparametrization".
The requirement that there are no other constants of motion except the energy seems to restrict its applicability to
one-dimensional systems. The theorem is mainly used 
to argue that symplectic integrators cannot be exact energy conserving. 
From the modern perspective of deriving SI from Lie operators and Lie series\cite{dep69,dra76,yos93,chin20},
rather than from generating functions\cite{fen10},
the theorem can be viewed as a corollary of approximating the exact evolution operator as a single product:
\be
\e^{\dt \hat H}=\e^{\dt(\hat T+\hat V)}\approx\prod_{i=1}^N \e^{a_i\dt \hat T}\e^{b_i\dt\hat V}=\e^{\dt \hat H ^\prime}
\la{evol}
\ee
where for the standard separable Hamiltonian $H=\p^2/2m+V(\br)$, $\hat T$ and $\hat V$ are Lie operators,
\be
\hat T=\bv\cdot\frac{\partial}{\partial \br}\qquad{\rm and}\qquad \hat V=
\bac(\br)\cdot\frac{\partial}{\partial \bv}
\la{tv}
\ee
with $\bv\equiv\p/m$, $\bac(\br)=-\nabla V/m$, and group actions
\ba
\e^{\dt \hat T}(\br,\bv)&=&(\br+\dt\bv,\bv)\nn\\
\e^{\dt \hat V}(\br,\bv)&=&(\br,\bv+\dt\bac).
\la{expbf}
\ea
The resulting modified Hamiltonian operator
\be
\hat H^\prime=\hat H+\dt \hat E_1+\dt^2 \hat E_2+\dt^3 \hat E_3+\cdots
\la{ham}
\ee
with various energy error terms $\hat E_k$, gives the trajectory of the algorithm via Hamilton's equation
with respect to the Hamiltonian function\cite{yos93} $H'$ corresponding to the operator $\hat H'$. 
If for some reasons $\hat H^\prime=\hat H$ with all $\hat E_k=0$,
then the trajectory produced by $H^\prime$ must be the same as the exact trajectory dictated by $H$.
From this perspective, exact energy conservation implies exact trajectory is a general feature of
symplectic integrators of the form (\ref{evol}), irrespective of whether there are other conserved quantities. 
This result only requires the existence of the modified Hamiltonian and its associated dynamics.

Moreover, the original proof of Ge\cite{ge88} allows for 
``time-reparametrization" of the form\cite{fen10} $H^\prime=F(\dt)H$, which usually does not occur
in practice. This is because $H^\prime$ as given by (\ref{ham}) must have
$F(\dt)=1$ due to the conventional requirements
\be
\sum_{i=1}^N a_i=1\quad{\rm and}\quad \sum_{i=1}^N b_i=1.
\ee
This can be easily verified
by applying the following second-order velocity-Verlet (VV) SI
\ba
\bv_1&=&\bv_0+\frac12 \dt \bac(\br_0)\la{vv1}\\
\br_1&=&\br_0+\dt \bv_1\la{vvp}\\
\bv_2&=&\bv_1+\frac12 \dt \bac(\br_1)
\la{vv}
\ea
corresponding to $N=2,a_1=0,b_1=1/2,a_2=1,b_2=1/2$ in (\ref{evol}),
and the position-Verlet (PV) SI
\ba
\br_1&=&\br_0+\frac12 \dt\bv_0\nn\\
\bv_1&=&\bv_0+\dt\bac(\br_1)\la{vpv}\\
\br_2&=&\br_1+\frac12 \dt\bv_1.
\la{pv}
\ea
corresponding to $N=2,a_1=1/2,b_1=1,a_2=1/2,b_2=0$ in (\ref{evol}),
to the case of a {\it constant} force with $\bac(\br)=\bac$. Both algorithms
then give the final position (\ref{vvp}) and (\ref{pv}) as 
$\br_{1,2}(\dt)=\br_0+\dt\bv_0+\frac12\dt^2 \bac$,
which is the exact trajectory for all $\dt$ with no time-reparametrization. 
The author is unaware of any example of the Ge-Marsden theorem 
with manifest time-reparametrization. It is therefore a complete surprise, that a 
highly non-trivial time-reparametrization is needed to obtain exact trajectories 
in a {\it constant} magnetic field. A resulting algorithm from such a time-reparametrization is 
none other than the well known Boris solver of plasma physics.

The Boris solver\cite{bun67,bor70,bir85} has been widely regarded as the
gold standard for solving charged particle trajectories 
in plasma\cite{bir85} and astrophysics\cite{rip18}. 
But why is it so good\cite{qin13}? 
Recent works focusing on volume-preserving methods\cite{qin13,he15} do not
explain why the Boris solver
is unique among all volume preserving algorithms. The real reason why the 
Boris solver is so good, according to Buneman\cite{bun67}, 
who proposed an earlier version of the solver, is that
``it leads to points on exact cycloids whenever E and B are constant". 
That is, for a constant magnetic field, 
positions generated by the Buneman solver\cite{bun67} 
are exactly on the gyro-circle, the cyclotron orbit. 
However, when Buneman's original $\E\times\B$ formulation was abandon\cite{bir85}
in favor of Boris's $\E$, $\B$ splitting scheme, this point  
seems to have been forgotten, especially when only the first-order leap-frog version of the
Boris solver is used\cite{bir85,par91,vu95,comm}. Otherwise, this on-orbit property
need not be pointed out again by others\cite{sto02} when using the 
second-order solver\cite{comm}.
 
The same operator method used to derive
symplectic integrator above has been used by the author to derive {\it exact energy conserving} (EEC) 
integrators for solving charged particle trajectories in a general magnetic field\cite{chin08}.
Since these algorithms use the Lorentz force law directly in terms of the 
{\it mechanical} momentum rather than the {\it canonical } momentum, these algorithms are
EEC but not symplectic. (They are referred to as {\it Poisson integrators} in Ref.\onlinecite{kna15}.) 
Because of this, their properties are not directly related to 
the Ge-Marsden theorem: they conserved the energy exactly, yet their trajectories are not exact. 
For a constant magnetic field, their orbits are not on the gyro-circle. 
According to Buneman\cite{bun67}, that can only be achieved with additional ``cycloid fitting". 
Remarkably, this work shows that this ``cycloid fitting" can be done exactly by a 
time-step reparametrization, a possibility suggested by the Ge-Marsden theorem.
One of the resulting algorithm is then the Boris solver.
 
This work presents a completely original derivation of Boris solvers  
independent of any finite-difference schemes, including a previously
unknown second Boris solver. To see how these two Boris solvers are related to
symplectic integrators, this work will review in the next section,
the magnetic velocity-Verlet and position-Verlet integrators
previously derived in Ref.\onlinecite{chin08}.
In Sect.\ref{fit}, the two Boris solvers 
are derived by ``cycloid fitting". Simple higher order algorithms are then
constructed in Sect.\ref{hi}. Conclusions are stated in Sect.\ref{con}.

\section {The two fundamental magnetic algorithms}
\la{two}

For a charged particle $q$ in a magnetic field ${\bf B(\br)}=B(\br)\bh(\br)$,
its acceleration is given by the Lorentz force law: 
\be
\frac{d\bv}{dt}=\w(\br)\bh(\br)\times\bv,
\la{velw}
\ee
where $\w(\br)=(-q)B(\br)/m$ is the local cyclotron angular frequency. 
(Note that $\w$ is correctly positive for $q<0$ because if $\bh(\br)$ is out of the page, 
the cyclotron motion is counterclockwise.) The operator $\hat V$ in (\ref{tv}) is then replaced by
\be
\hat V_{B}
=\w(\bh\times\bv)\cdot\frac{\partial}{\partial \bv},
\ee
whose effect on $\bv$ is just
\ba
\hat V_{B}\bv&=&(\w\bh\times )\bv\nn\\
\hat V_{B}^n\bv&=&(\w\bh\times )^n\bv
\ea
and therefore the velocity update is
\ba
\bv_B(\br,\bv,\dt)&=&\e^{\dt \hat V_B}\bv=\e^{\theta(\bh\times\ )}\bv\la{expc}\\
&=&\bv+(\ta-\frac{\ta^3}{3!}+\frac{\ta^5}{5!}+\cdots)\bh\times\bv
+(\frac{\ta^2}2-\frac{\ta^4}{4!}+\cdots)\bh\times(\bh\times\bv)\la{rsum}\\
&=&\bv+\sin\ta\bh\times\bv+(1-\cos\ta)\bh\times(\bh\times\bv)\la{vrot}
\ea
where $\theta=\w(\br) \dt$ and where one has repeatedly used the identity
\be
\bh\times(\bh\times(\bh\times\bv))=-\bh\times\bv
\ee
to reduce the exponential series to (\ref{rsum}) and resummed as (\ref{vrot}), which is the exact 
solution to (\ref{velw}) holding $\br$ fixed.
By decomposing $\bv$ into components parallel and perpendicular to $\bh(\br)$
\be
\bv=\bv_{||}+\bv_{\perp}, 
\ee
so that $\bh\times(\bh\times\bv_{\perp})=-\bv_{\perp}$, (\ref{vrot}) becomes 
\be
\bv_B(\br,\bv,\dt)=\bv_{||}+\cos\ta\bv_{\perp}+\sin\ta(\bh\times\bv_{\perp}).
\la{exrot}
\ee
This shows that the magnetic field only rotates the perpendicular velocity component
without changing the velocity's magnitude and hence the kinetic energy.
 
By replacing the velocity updates in (\ref{vv1}), (\ref{vv}) and (\ref{vpv}) by (\ref{vrot}),
or (\ref{exrot}), one obtains the corresponding magnetic integrator M2a 
\ba
\bv_1&=&\bv_B(\br_0,\bv_0,\dt/2)\la{vvf}\\
\br_1&=&\br_0+\dt \bv_1\la{mvv}\\
\bv_2&=&\bv_B(\br_1,\bv_1,\dt/2)
\la{vvl}
\ea
and M2b
\ba
\br_1&=&\br_0+\frac12\dt \bv_0\nn\\
\bv_1&=&\bv_B(\br_1,\bv_0,\dt)\la{mpv}\\
\br_2&=&\br_1+\frac12 \dt \bv_1
\la{pvr}
\ea
as derived in Ref.\onlinecite{chin08}. Each is completely explicit, sequential, exact energy conserving,
but {\it not} symplectic. This is because in order to be symplectic, one must
update the canonical momentum in Hamilton's equations of motion. Since the Lorentz force law (\ref{velw})
uses only the mechanical momentum, any direct use of the Lorentz force law is non-canonical and
therefore non-symplectic.

To see how these magnetic integrators work, consider the case of an electron in a constant magnetic field
with $\w=2$, $\bh=\hat {\bf z}$, $\br=(x,y)$, $\bv=\bv_\perp=(v_x,v_y)$,
with initial velocity $\bv_0=(0,v_0)$, $\br_0=(r_g,0)$, where $v_0=1$ and where
$r_g=v_0/\w=1/2$ is the gyro-radius.
Let's take a large $\dt=\pi/4$, so that $\ta=\w\dt=\pi/2$. 
Therefore both $\br$ and $\bv$ rotate through $90^\circ$ 4 times to complete one orbit. 
The M2a and M2b trajectories are the square orbits shown in Fig.1. 
For M2a, the position update (\ref{mvv}) is 
$\dt$ times the initial velocity rotated $45^\circ$, so the length of the displacement 
is $v_0\dt$. For M2b, the position update
(\ref{pvr}) is $\dt$ times half the diagonal of a square with sides $v_0$, 
which is $v_0\dt/\sqrt{2}$. Therefore the displacement of M2a is $\sqrt{2}$ times that of M2b, 
as shown. Their trajectories fall outside and inside of the gyro-circle respectively.

\begin{figure}[hbt]
	\includegraphics[width=0.98\linewidth]{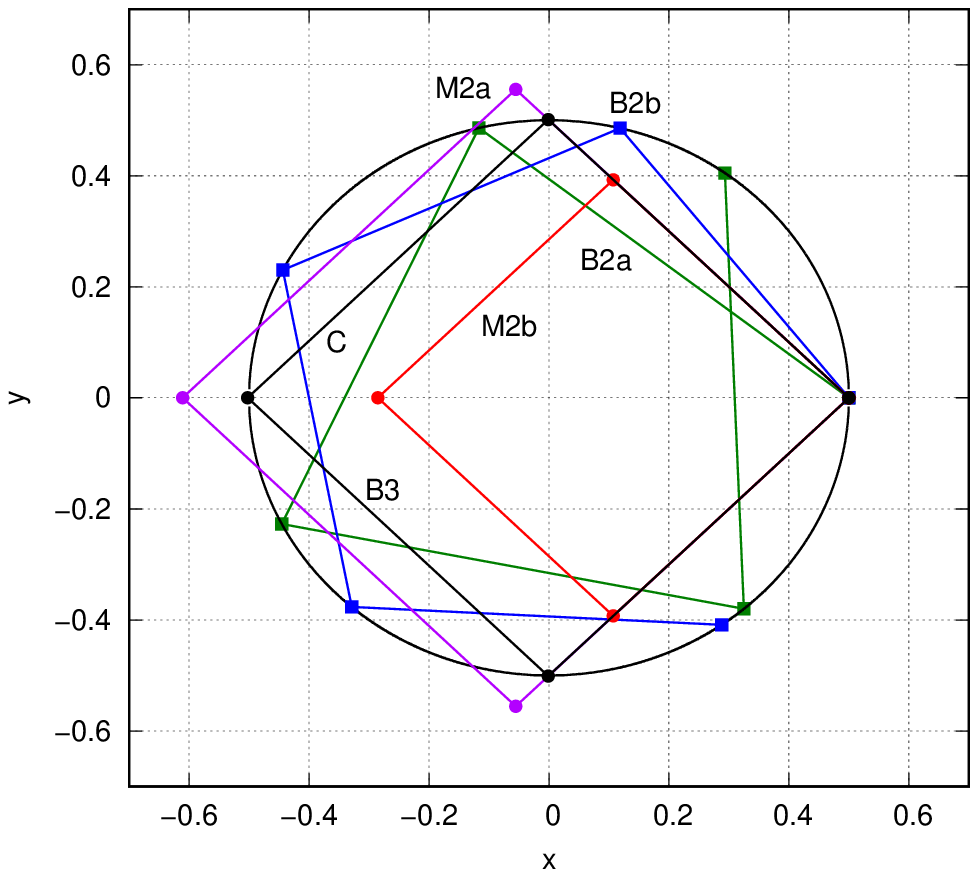}
	\caption{ (color online)
		All orbits start on the rightmost point on the gyro-circle. 
		M2a and M2b have three points outside and inside of the circle respectively. 
		B2b and B2a have points on the circle but with angles behind and ahead of the phase correct
		angle of M2a and M2b. C and B3 are higher order results having very small error off
		the circle and very small phase error respectively. See text for details. 		
	}
	\la{algab}
\end{figure}

\section{cycloid fitting}
\la{fit}

For M2b, the position updating step (\ref{pvr}) is
\ba
\br_2&=&\br_0+\frac12\dt(\bv_0+\bv_1)\nn\\
&=&\br_0+\frac12\dt(\bv_0+\cos\ta\bv_0+\sin\ta\bh\times\bv_0)
\la{rsq}
\ea 
Defining $\br_g=\bv_0/\w=(0,r_g)$ and noting that $\br_0=(r_g,0)=-\bh\times\br_g$, one can rewrite the
above as
\ba
\br_2
&=&\frac\ta{2}(1+\cos\ta)\br_g-(1-\frac\ta{2}\sin\ta)\bh\times\br_g),\la{rup}
\ea
and therefore the trajectory is always inside of the gyro-circle:
\ba
|\br_2|^2
&=& r_g^2\left[1-2\ta\cos^2(\frac{\ta}{2})\Bigl(\tan(\frac{\ta}{2})-\frac{\ta}{2}\Bigr)\right]\la{bor1}\\
&=& r_g^2[1-\frac1{12}\ta^4+\frac1{80}\ta^6+\cdots]\ {\rm as}\ \ta\rightarrow 0.
\la{rgm}
\la{rr}
\ea 
However, if one were to {\it decouple} the actual rotation angle (arguments of trig functions) 
from $\ta$ to an effective angle $\ta_B$, then requiring 
\be
\tan(\frac{\ta_B}{2})=\frac{\ta}{2}
\la{bcon}
\ee
in (\ref{bor1}) would force the trajectory to be exactly on the gyro-circle!
This corresponds to a highly non-trivial reparametrization of $\dt$ in the rotation angle 
to a Boris time step $\dt_B$ via
\be
\dt_B=\frac2{\w}\tan^{-1}(\w\dt/2).
\ee
If the updating rotation step (\ref{exrot}) is replaced by
\be
\bv_B(\br,\bv,\dt)=\bv_{||}+\cos\ta_B\bv_{\perp}+\sin\ta_B(\bh\times\bv_{\perp}),
\la{brot}
\ee
then M2b reproduces the Boris solver denoted here as B2b.
From (\ref{bcon}), $\cos\ta_B$ and $\sin\ta_B$ can be easily reconstructed as
\ba
\sin\ta_B&=&\frac{2\tan(\ta_B/2)}{1+\tan^2(\ta_B/2)}
= \frac{\ta}{1+\ta^2/4},\la{tsin}\\
\cos\ta_B&=&\frac{1-\tan^2(\ta_B/2)}{1+\tan^2(\ta_B/2)}
= \frac{1-\ta^2/4}{1+\ta^2/4}.
\la{tcos}
\ea
This makes it clear that the position vector (\ref{rup})
\ba
\br_2
&=&\frac{\ta}{1+\frac14\ta^2}\br_g
	-\frac{1-\frac14\ta^2}{1+\frac14\ta^2}\bh\times\br_g\la{br2}\\
&=&\sin(\ta_B)\br_g-\cos(\ta_B)\bh\times\br_g\nn\\
&=&\cos(\ta_B-\pi/2)\br_g+\sin(\ta_B-\pi/2)\bh\times\br_g,
\la{bre}
\ea
is not only precisely on the gyro-circle, but it lags $90^\circ$ behind, and is therefore
perpendicular to the Boris-angle rotated velocity vector (\ref{brot}).

The resulting trajectory in Fig.1 is labeled as B2b. Notice that B2b's angle of rotation lags behind that of the M2a and M2b. This follows from $\ta=\pi/2=1.5708$,
\be
 \ta_B=2 \tan^{-1}(\ta/2)=1.3316\quad{\rm and}\quad \ta_B-\ta=-0.2392\ .
\la{lag}
\ee
More generally at small $\ta$, the Boris angle $\ta_B$ has a second order phase error,
\be
\ta_B= \ta[1 -\frac1{12}\ta^2+\frac1{80}\ta^4+\cdots],
\la{rot1}
\ee
showing B2b's rotation angle always lags, but remarkably with the same two leading 
coefficients as in gyro-radius (\ref{rgm}). For extensive comparisons of B2b with M2a and M2b 
on {\it nonuniform} magnetic fields, see the work by Knapp, Kendl, Koskela and Ostermann\cite{kna15}.

Since $\ta_B$ can be defined via (\ref{bcon}) for all $\dt$, B2b's trajectory will stay on the gyro-circle 
for all $\dt$, no matter how large. In the limit of $\ta\rightarrow\infty$,
$\ta_B\rightarrow\pi$, B2b's
trajectory will just bounce back and forth nearly as straight lines across the diameter of the gyro-circle. 
Nevertheless, this means that B2b's trajectory remains bounded even as $\dt\rightarrow\infty$!
It is unusual that an explicit algorithm can be unconditionally stable. 

Conventionally, the stability of the original Boris solver was thought to be connected to 
the Cayley approximation for the exponential operator in (\ref{expc}): 
\be
\e^{\theta(\bh\times\ )}\approx 
[1-\frac12\theta(\bh\times\ )]^{-1}[1+\frac12\theta(\bh\times\ )]\equiv {\bf C},
\la{cay}
\ee
found in the implicit midpoint approximation used in Boris's original derivation\cite{bor70}.
By expanding the inverse operator as $1/(1-x)=1+x+x^2+\cdots$ and recollecting terms as
in (\ref{rsum}), one finds the same Boris rotation:
\ba
{\bf C}\bv
&=&[1+\frac{\frac12\ta \cc+\frac14\ta^2\cc^2}{1+\frac14\ta^2}][1+\frac12\ta(\bh\times\ )]\bv,\nn\\
&=&[1+\frac{\ta \cc+\frac12\ta^2\cc^2}{1+\frac14\ta^2}]\bv,\nn\\
&=&\bv_{||}+\frac{1-\frac14\ta^2}{1+\frac14\ta^2}\bv_{\perp}+	 
\frac{\ta}{1+\frac14\ta^2} (\bh\times\bv_{\perp}).
\la{cal}
\ea
However, by itself, this Boris rotation (\ref{cal}), no more than the exact rotation (\ref{exrot}),
can confer greater stability. That is, if this Boris rotation (\ref{cal}) is just another {\it mathematical}
means of rotating the velocity vector (but without evaluating trigonometric functions), then there is 
no {\it a prior} reason to expect that it would work better than the exact rotation (\ref{exrot}).
However, this derivation showed that the Boris angle condition (\ref{bcon}) forces
the trajectory to be on the gyro-circle. It is by enforcing this {\it physical} property
that the Boris rotation can achieve greater, and in this case, unconditional stability.
In principle, the angle condition (\ref{bcon}) is totally sufficient to define the Boris rotation
(\ref{brot}) via (\ref{tsin}) and (\ref{tcos}). It is entirely irrelevant whether this
angle condition can be realized by any specific approximation 
of the exponential operator such as the Cayley transform (\ref{cay}).

For M2a, the position updating step (\ref{mvv}) is 
\ba
\br_1&=&\br_0+\dt[\cos(\ta/2)\bv_0+\sin(\ta/2)\bh\times\bv_0]\la{vvs}\\
&=&\ta\cos(\ta/2)\br_g+(\ta\sin(\ta/2)-1)\bh\times\br_g)\la{rpp}\\
|\br_1|^2 &=&r_g^2\left[1+2\ta \left( \frac{\ta}{2}-\sin(\frac{\ta}{2} ) \right)   \right]\nn\\
&=& r_g^2[1+\frac1{24}\ta^4-\frac1{1920}\ta^6+\cdots]\ {\rm as}\ \ta\rightarrow 0.
\la{rgp}
\ea
Therefore, if one now defines a second Boris angle $\ta'_B$ by
\be
\sin(\frac{\ta'_B}{2} )=\frac{\ta}{2},
\la{svv}
\ee
and set
\be 
\cos (\frac{\ta'_B}{2} )=\sqrt{1-(\frac{\ta}{2})^2}
\ee
in (\ref{vvf}) and (\ref{vvl}), then one obtains a second, previously unknown, 
Boris solver whose trajectory is
also exactly on the gyro-circle. The position vector (\ref{rpp}) is then
\ba
\br_1&=&\ta\sqrt{1-(\frac{\ta}{2})^2}\ \br_g-(1-\ta^2/2)\bh\times\br_g)\la{rp2}\\
     &=&\sin(\ta'_B)\br_g -\cos(\ta'_B)\bh\times\br_g\nn\\
     &=&\cos(\ta'_B-\pi/2)\br_g+\sin(\ta'_B-\pi/2)\bh\times\br_g,
\ea
which is again $90^\circ$ behind the $\ta'_B$ rotated velocity.

This second Boris solver's orbit is labeled as B2a in Fig.1. Here, $\ta'_B$ is ahead of M2a and M2b by an amount nearly equal to that of (\ref{lag}):
\be
\ta'_B=2 \sin^{-1}(\ta/2)=1.8067\quad{\rm and}\quad \ta'_B-\ta=0.2359\ .
\la{head}
\ee
At small $\ta$, $\ta'_B$ remains ahead, with second-order error 
\be
\ta'_B=\ta[1 +\frac1{24}\ta^2+\frac3{640}\ta^4+\cdots]
\la{rot2}
\ee
having the same leading coefficient as in (\ref{rgp}), but not the exact opposite of (\ref{rot1}).

From (\ref{svv}), this $\ta'_B$ can only be defined for $|\ta/2| \le1$. Therefore this
second Boris solver is limited to $|\dt|<|2/\w|=T/\pi$, approximately a third of the gyro-period.
This same range of stability is obtained when the symplectic VV integrator is applied to the harmonic oscillator\cite{chin20}. 

This second Boris solver also underscores the fact it is the second Boris angle condition (\ref{svv}) that is of
paramount importance. The use of the Cayley transform (\ref{cay}) here would not 
have improved the M2a integrator in any way. 
Also, it is not clear whether this second Boris condition (\ref{svv}) can be realized by 
a Cayley-like operator approximation of the exponential, as in (\ref{cay}). 

Finally, note that for motion perpendicular to $\bh$, since
\be
\frac{d}{dt}(\br^2)=2\br\cdot\bv\quad{\rm and }\quad \frac{d(\br\times\bv)}{dt}=\w(\br\cdot\bv)\bh,
\la{rsq2}
\ee
the constancy of $r^2$, the orthogonality of $\bv$ to $\br$, and
the conservation of angular momentum, are one and the same condition.

\section{Higher order algorithms}
\la{hi}

Given basic second-order algorithms M2a, M2b and B2b, B2a, well known methods\cite{chin08,he15,kna15} can be
used to generate higher order schemes. However, some schemes are so simple that they can be
constructed by inspection. For example,
in view of (\ref{rgm}) and (\ref{rgp}), the combination algorithm
\be
C(\dt)= \frac23 M2a(\dt)+\frac13 M2b(\dt),
\ee
would have sixth-order $r_g^2$ error at small $\dt$:
\be
|\br|^2=r_g^2[1+\frac{11}{2880}\ta^6+\cdots].
\ee
Surprisingly, it is already very good even at $\dt=\pi/4$. This is shown as the square orbit $C$ in Fig.1. 
As pointed out previously, the length of the M2a step is $v_0\dt=\pi/4$, the length of the M2b step is 
$(\pi/4)/\sqrt{2}$. The length of the combination C step is therefore
\be 
 \frac23 \frac{\pi}4 +\frac13 \frac{\pi}4\frac1{\sqrt{2}}=\frac\pi{12}(2\sqrt{2}+1)
 \frac1{\sqrt{2}}\approx 1.002 \frac1{\sqrt{2}}
\ee
Since the length require to be on the gyro-circle is $1/\sqrt{2}$,
algorithm C's slight over-shoot is not visually visible. 
Since C has no phase error, it is very close to, but not exactly on the gyro-circle.

For B2b and B2a, since their trajectories are already on the gyro-circle, one can only improve
their phase error. To preserve their positions on the gyro-circle, one must apply B2b and B2a sequentially.
In view of (\ref{rot1}) and (\ref{rot2}), a time-symmetric sequence is
\be
B3(\dt)=B2a(\frac{\dt}3 )B2b(\frac{\dt}3 )B2a(\frac{\dt}3 )
\ee
with Boris angle
\ba
\ta_{B3}&=&4\sin^{-1}(\ta/6)+2 \tan^{-1}(\ta/6)\nn\\
&&\qquad= \ta [1+\frac7{77760}\ta^4 \cdots].
\la{tb3}
\ea
For the case considered, $\ta=\pi/2$,
$$\ta_{B3}=1.0005(\frac\pi{2}),
$$
which is, surprisingly, already in agreement with the small angle estimate (\ref{tb3}). 
The resulting orbit, labeled B3, overlaps with orbit C in Fig.1.

\section{Conclusions and future directions}
\la{con}

In this work, two Boris solvers are derived from two fundamental magnetic field integrators M2b and M2a. 
This derivation demonstrated the underlying unity among all second-order explicit algorithms, 
from symplectic, magnetic to Boris integrators. There is no need to tinker with various 
finite-difference schemes or concoct {\it ad hoc} adjustments. 
There are only two fundamental algorithms in all cases, 
from which higher order schemes can be systematically constructed.

As shown in the Introduction, a constant force, including that from a constant electric field,
can be exactly integrated by a second-order algorithm. Therefore, for a constant electric and magnetic
field, cycloid fitting, as shown by Buneman solver\cite{bun67}, remains just fitting the gyro-circle.
The added electric field can be easily accounted for by splitting\cite{bor70,kna15}. 

Boris originally suggested\cite{bor70,sto02} the $\tan(\alpha)/\alpha$ modification,
which is to replace $\ta/2$ by $\tan(\ta/2)$, so that his solver's phase would be 
correct by retracing (\ref{tsin}) and (\ref{tcos}) in the reverse order. 
However, this work showed that such a modification would simply revert B2b back to M2b, 
losing large $\dt$ stability and going off the gyro-circle.

From (\ref{expbf}), the characteristics of symplectic integrators are
position and velocity translations. From (\ref{exrot}), the characteristics of magnetic
integrators are position translations and velocity rotations. 
They are therefore fundamentally different. 
For symplectic integrators, the Ge-Marsden theorem correctly single out the algorithm's energy,
the modified Hamiltonian, as the controlling factor in determining 
the trajectory's accuracy. For purely magnetic integrators, because they are not
Hamiltonian based, the controlling factor is not known. It is not clear what plays
the role equivalent to the modified Hamiltonian.  
This work showed that exact energy conserving magnetic integrators must have reparametrized 
time steps, or phase errors, in order to produce exact trajectories. Yet, the derivation is 
just retrofitting, one has no theory of how to exploit the phase freedom to achieve greater 
trajectory accuracy. All these suggest that an equivalent Ge-Marsden theorem for magnetic 
integrators would be most helpful in guiding the future development of more efficient 
magnetic algorithms.    

%



\begin{thebibliography}{10}
\bibitem{comm} The confusion and distinction between the first and second-order Boris 
solvers will be clarified in a forthcoming article.
\end{thebibliography}
\end{document}